\documentclass[conference]{IEEEtran}

\usepackage[pdftex]{graphicx}
\usepackage{amsmath}
\usepackage{eqparbox}
\usepackage{xcolor}
\usepackage{caption}
\usepackage{subcaption}
\usepackage{amssymb}
\usepackage{mathtools}
\usepackage[utf8]{inputenc}
\usepackage{float}
\usepackage{booktabs}
\usepackage{fancyhdr}
\begin{document}
%+++++++++++++++++++++++++++++++++++++++++++++++++++
\title{\LARGE Doppler Invariant Demodulation for Shallow Water Acoustic Communications Using Deep Belief Networks}
\author{\authorblockN{Abigail Lee-Leon\authorrefmark{1}\authorrefmark{2}, Chau Yuen\authorrefmark{1}, Dorien Herremans\authorrefmark{1}}
\authorblockA{\authorrefmark{1}Singapore University of Technology and Design (SUTD), 8 Somapah Road, Singapore 487372}
\authorblockA{\authorrefmark{2}Thales Solutions Asia Pte Ltd, 21 Changi North Rise, Singapore 498788}}
%+++++++++++++++++++++++++++++++++++++++++++++++++++
\maketitle
% ================
% # Abstract     #
% ================

\begin{abstract}
Shallow water environments create a challenging channel for communications. In this paper, we focus on the challenges posed by the frequency-selective signal distortion called the Doppler effect. We explore the design and performance of machine learning (ML) based demodulation methods --- (1) Deep Belief Network-feed forward Neural Network (DBN-NN) and (2) Deep Belief Network-Convolutional Neural Network (DBN-CNN) in the physical layer of Shallow Water Acoustic Communication (SWAC). The proposed method comprises of a ML based feature extraction method and classification technique. First, the feature extraction converts the received signals to feature images. Next, the classification model correlates the images to a corresponding binary representative. An analysis of the ML based proposed demodulation shows that despite the presence of instantaneous frequencies, the performance of the algorithm shows an invariance with a small 2dB error margin in terms of bit error rate (BER).
\end{abstract}


\pagestyle{fancy}
\renewcommand{\headrulewidth}{0pt}
\fancyhf{}

\fancypagestyle{plain}{%
\renewcommand{\headrulewidth}{0pt}%
\fancyhf{}%
\fancyfoot[L]{\footnotesize \center \emph{Preprint accepted for publication in the 16th IEEE Asia Pacific Wireless Communications Symposium (APWCS). Singapore, 2019, \\ -\thepage-}}%
}

\cfoot{\footnotesize \center \emph{Preprint accepted for publication in the 16th IEEE Asia Pacific Wireless Communications Symposium (APWCS). Singapore, 2019, \\ -\thepage-}}  

% ========================
% # I. Introduction      #
% ========================

\section{Introduction}

Shallow Water Acoustic Communications (SWAC) channels are generally recognized as one of the most difficult communication media in use today. One of the main reasons for this is the change in instantaneous frequencies after propagation called the Doppler effect. The severity of which is amplified for SWAC channels due to the relatively slow speed of sound~\cite{A.C.Singer}. It is caused both by the delays spreading over tens or hundreds of milliseconds~\cite{K.C.H.} and the relative displacement of the transmitter and receiver. The challenge of the Doppler effect in SWAC channel is especially complex and requires more effective methods to handle the problem. 

When the challenge is to model complex system, machine learning (ML) or deep learning (DL) is the technique of choice for most researchers~\cite{Zheng2010}. There has been growing interest in applying DL in fields like image recognition~\cite{Mashford} and natural language processing~\cite{Young}. Recent works done by Wang et al.~\cite{Wang} exploited deep learning to detect signal modulations in underwater channels. Applying ML techniques to communication blocks has provided a promising solution to the complex channel problem. 

Demodulation is seen as one of the most fundamental blocks in communications systems. It works by using an estimation of the information stored in the instantaneous phase signals, given the composite signal. The conventional method of demodulation is called the maximum-likelihood estimation (MLE), which employs an iterative approach based on Newton's algorithm. It uses the discrete-time polynomial phase transformation~\cite{Friedlander} to differentiate the binary representatives of the received signals, resulting in demodulation. This method is dependent on the knowledge of the modulated signal's carrier frequency. Therefore, the changes in instantaneous frequency, by the Doppler effect due to the SWAC channel, greatly diminishes the accuracy of the conventional demodulation technique. 

To address this problem, we propose two novel approaches to the demodulation problem --- (1) Deep Belief Network-feed forward Neural Network (DBN-NN) and (2) Deep Belief Network-Convolutional Neural Network (DBN-CNN). The proposed architecture consists of a feature extraction technique and followed by a classification method. The feature extraction method comprises of a Deep Belief Network (DBN). It constitutes multiple Restricted Boltzmann Machines (RBMs) stacked together to form a network that learns distinctive features from the received signal and translates these features into an image matrix. This feature image is inputted to the classification models for labelling. The classification model uses NNs to associates the image with its binary representation, resulting in signal demodulation.

The remainder of this paper is organized as follows. Section~\ref{Communication} describes the communication system. Section~\ref{Proposed methods} illustrates the feature extraction method and the two classifying architectures, feed forward NN and CNN. Section~\ref{discuss} provided and discussed the simulation results. Finally, conclusions are drawn in Section~\ref{conclusion}.

% ====================
% # II. System Model #
% ====================

\section{End-to-end Communication System}
\label{Communication}

In SWAC, the channel has been known to be too complex to understand and model~\cite{Stojanovic.M}. One of the major challenges is the varying instantaneous frequency called the Doppler effect. Unlike in wireless communications, the Doppler effect is prevalent in the underwater channel. It is caused by scatter of the signals due to environmental changes, propagation over several paths and long delays caused by the underwater speed of sound~\cite{Stojanovic1994}.

\begin{figure}[h] %!t
\centering
\includegraphics[width=8.5cm]{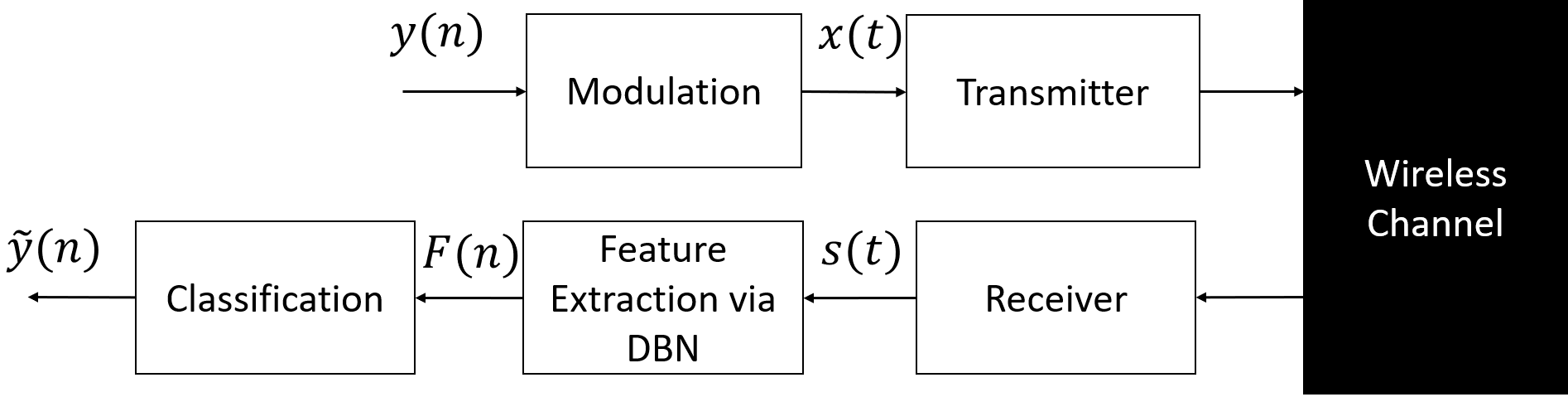}
\caption{End-to-end System Model}
\label{System Model}
\end{figure}

First, we consider the proposed overall communication system represented by Fig.~\ref{System Model}. The system consists of a single transmitter and receiver.

In this paper, we focus on the changes in instantaneous frequency in the SWAC channel called the Doppler effect. Let $y(n)$ be the binary representation of the transmitted signal $x(t)$ during the \textit{n}-th transmission. A series of transmission symbols $y(n)$ are translated into different transmission signal waveforms $s(t)$ via a Phase Shift Key (PSK) modulator. The following equation is used for the modulation.
\begin{equation}
    x(t) = cos(2 \pi f_c t + \theta_{n})
    \label{BPSK}
\end{equation}
for the period $0 \leq t \leq T$, where T is the duration of a bit, $\theta_n$ is the phase of the signal and $f_c$ is the carrier frequency.

Secondly, we simplify the channel complexity by focusing on the Doppler effect and noise. $x(t)$ is relayed through the following channel model.
\begin{equation}
s(t) = x(\alpha_i t) + n(t)
\label{channel}
\end{equation}
where $\alpha_i$ represents the Doppler scaling factor and $n(t)$ is the Addictive White Gaussian Noise (AWGN).

Finally, we assume a set of training signals $L = {(x_t, y_t)}$, $t = 1, 2, ..., n$, where $x_t$ is a training signal, $y_t$ is the corresponding label vector and $n$ is the number of the training signals. The objective of the proposed algorithm is to build a model from the training data $L$, such that for a given test signal $x(t)$, the learned model will be able to construct a predicted label $\Tilde{y}(n)$. 

To classify the signals, we consider the overall problem of estimating the labels $\Tilde{y}(n)$ via a learning function $f(\cdot)$. The following equation provides a mathematical interpretation of the problem.
\begin{equation}
    \Tilde{y}(n) = f(s(t))
    \label{Problem}
\end{equation}

% ====================================
% # III. Feature Extraction via DBN #
% ====================================
\section{Proposed Methods}
\label{Proposed methods}

In this section, the proposed methods (1) DBN-NN and (2) DBN-CNN are described. First, the feature extraction pre-processing of the received signal into a feature image is accomplished via DBN. After which, the feature images are matched to their corresponding binary representatives using a classifier. 

\subsection{Feature Extraction: DBN}
\label{Feature Extraction via DBN}

For the feature extraction pre-processing, the input is the framed received signal $s(t)$, that is segregated into vectors of length (1$\times$120) with a 20\% overlap. For the pre-training of the feature extraction, the corresponding labels used to check the performance of the feature extraction are the binary representatives. For example, the bit 0 translated into its modulated waveform $s_0(t)$ would have the label of 0 during the pre-training check. 

DBN is a learning framework based on a deep NN with an unsupervised pre-learning ability~\cite{Salama}.  It comprises of multiple RBMs stacked together to execute layer-wise greedy and unsupervised learning. Every layer of DBN extracts features and translates the layer's input to a more conceptual representation.

RBM is created using probabilistic binary units that work in a stochastic manner. It is comprised of a two layer NN --- the visible layer $v$ and the hidden layer $h$. For RBM, every node in one layer is connected to all nodes of the next layer. However, these connections are bi-directional and there are no links between nodes of the same layer. 

The energy of the joint configuration in Boltzmann machines is given as follows:
\begin{equation}
    E(v,h) = -\sum_{k=1}^{u_v} \sum_{j=1}^{u_h} hWv-\sum_{k=1}^{u_v}bv-\sum_{j=1}^{u_h}ch
    \label{Energy}
\end{equation}
where the visible nodes $v \in \mathbb{R}$ correspond to the input and $u_v$ is the number of visible nodes, the hidden nodes $h \in \mathbb{R}$ represents the latent features and $u_h$ is the number of hidden nodes, $W$ is the concurrent weights linking the nodes of the visible to hidden layer, $c$ and $b$ are the bias terms of the hidden and visible nodes respectively.

$v$ and $h$ are assigned an energy probability value that is defined as:
\begin{equation}
    p(v,h) = \frac{1}{Z} \exp{(-E(v,h))}
    \label{Energy partition}
\end{equation}
where $Z$ is the partition function that is obtained via:
\begin{equation}
    Z = \sum_{k=1}^{u_v} \sum_{j=1}^{u_h} \exp{(-E(v,h))}
    \label{Energy partition}
\end{equation}

To optimize the parameters of the network at each layer $k$, the following optimization problem shown by Eqn.~\ref{Objective} is minimize via partial differentiation with respects to $W,b,c$.
\begin{equation}
    g_k(v,h) = - \frac{1}{m} \sum_{j=1}^{m}\log(P(v_k^j,h_k^j))
    \label{Objective}
\end{equation}
% ==========================
% # IV. CNN based Demodulation #
% ==========================

\subsection{Classification Architectures}
\label{Classification A.}
In the following subsections, we will use two different NN architectures to classify the feature image produced by the DBN --- (1) Feed forward NN and (2) CNN.~\\
\subsubsection{Feed forward NN}

Feed forward NN are used for classification and regression. The network works as a classifier to map the input $F(n)$ to the label $y(n)$ via Eqn.~\ref{NN}.
\begin{equation}
y(n) = g(F(n);\phi)
\label{NN}
\end{equation}
where $\phi$ represents the NN parameters --- weights, $W_\phi$, and bias, $b_\phi$.

The NN architecture consists of 3 layers. The parameters of the layers are shown in Table \ref{tab:my-table1}. 
\begin{table}[b]
\centering
\caption{Feed forward NN Layers using sigmoid activation}
\label{tab:my-table1}
\scalebox{1}{%
\begin{tabular}{ccc}
\toprule
Layers & Input size & Output size\\ \midrule
1 &  784 & 300\\ 
2 &  300 & 50 \\ 
3 &  50 & 2  \\ \bottomrule
\end{tabular}}
\end{table}
$F(n)$ is reshaped into a (781$\times$1) matrix and used as the input of this network. To optimize the network, the cross entropy loss function shown in Eqn.~\ref{loss} is used. Cross entropy measures the degree of separation between two probability distributions. The objective is to minimize the cross entropy, such that the two distributions are seen as similar.
\begin{equation}
L = -\frac{1}{n} \sum_{m=1}^{n} [y(m)\log(\Tilde{y}(m))+(1-y(m))\log(1-\Tilde{y}(m)]
\label{loss}
\end{equation}

As such, the objective is to decrease the loss function using gradient descent depicted by Eqn.~\ref{gradient}.
\begin{equation}
\min \bigg\{\frac{dL}{d\phi} = (y(n)-\sigma(W_\phi s(t)+b_\phi))\cdot s(t)\bigg\}
\label{gradient}
\end{equation}
\subsubsection{CNN}

\begin{figure}[h]
\centering
\includegraphics[width=9cm]{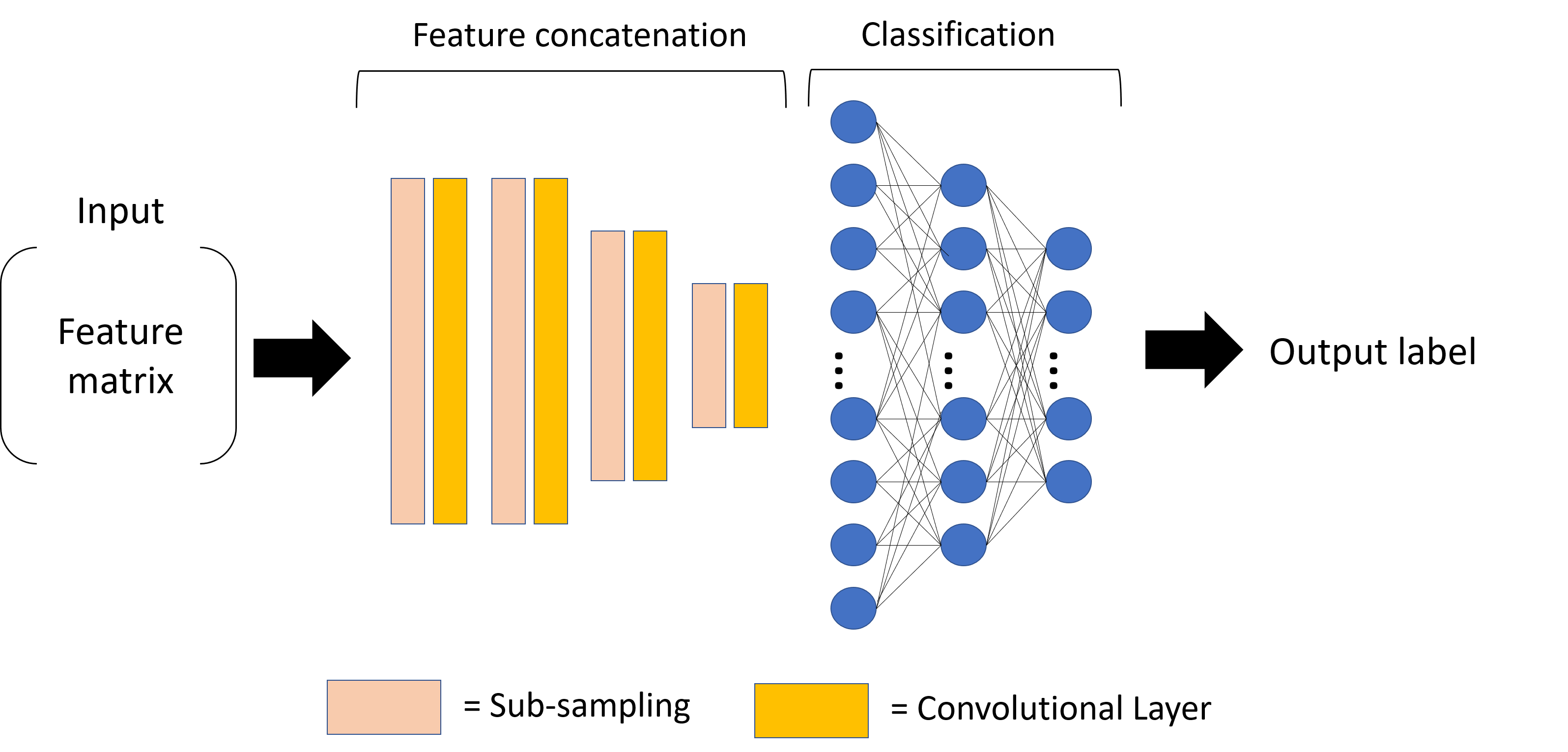}
\caption{CNN Architecture using alternating convolutional and sub-sampling layers with a NN classification}
\label{Overview of CNN model}
\end{figure}

The architecture of our CNN is shown in Fig.~\ref{Overview of CNN model}. There are 3 main parts in the model --- a convolutional, down-sampling and fully connected layers. 

The convolutional layer is defined by the size of the convolutional kernels. The size of the kernel (n$\times $n) is typically much smaller than the (m $\times$ m) input feature matrix $F(n)$, as shown in Table.\ref{CNN layers}. These kernels are linked to small areas of the image known as a receptive field. They slide horizontally and vertically to extract different sets of features by convolving its weights with the image area to produce $k$ feature maps of size (m-n+1$\times$m-n+1). 
These feature maps are then sub-sampled with max pooling over non-overlapping rectangular regions of size (5$\times$5). Max-pooling enables position invariance over larger local regions and down-samples the input image by a factor of 5 along each direction. Max-pooling improves performance by selecting dominant homogeneous features using a rapid convergence rate. 

The final layers of the network consist of two fully connected layers with an output node corresponding to each classification label. The softmax activation function shown in Eqn.~\ref{softmax} is used to calculate the probabilities of on instance belonging to each class $y_n$. 
\begin{equation}
P(s_i(t) = y_i(n) \mid x_i(t)) = \frac{e^{(x_i(t)  W_{y})}}{\sum_{m=1}^{M}e^{(x_i(t) W_m)}}
\label{softmax}
\end{equation}
where $s_i(t)$ is the framed received signal, $y_i(n)$ is the corresponding label, $x_i(n)$ is the corresponding transmitted signal, $W_m$ is the weights of the $m$-dimensional vector $x(t)  W_{y}$ and $W_{y}$ is the weight of each label $y_n$. 
\begin{table}[h]
\centering
\caption{CNN model has 7 layers. The S, P and Nodes denotes the sub-sampling, the size and number of the kernel (width $\times$ height $\times$ no of feature maps) and the number of output nodes.}
\label{tab:my-table}
\scalebox{1}{%
\begin{tabular}{cc}
\toprule
Input: & $F(n)$, (28$\times$28)\\ \midrule
Layer 1 & Convolutional, P = ($5\times5\times6$)  \\ 
Layer 2 & Pooling, S=2  \\ 
Layer 3 & Convolutional, P = ($5\times5\times16$)  \\ 
Layer 4 & Pooling, S=2 \\ 
Layer 5 & Convolutional, P = ($5\times5\times120$)  \\ 
Layer 6 & Neural Network, Nodes = 84\\ 
Layer 7 & Neural Network, Nodes = 2 \\  \midrule
Output & Predicted Label\\ \bottomrule
\label{CNN layers}
\end{tabular}}
\end{table}

% ========================
% # IV. Experimental Results and Discussion #
% ========================

\section{Results and Discussion}
\label{discuss}

In this section, we evaluate the DBN based feature extraction. After which a comparison of the overall proposed techniques --- (1) DBN-NN and (2) DBN-CNN and training strategies is completed. As a comparison, we used the conventional MLE method devised in~\cite{Stocia1989}.

For the following simulation experiments, the simulated dataset contains 40,000 transmitted signals periods, in which 50\% is used for training, 20\% on validation and the remaining 30\% on testing. The channel used for each experiment will be explained below. 

\subsection{Signal Feature Extraction}
\label{Sign Feat}

To evaluate the performance of the DBN feature extraction, the following experiments were concluded. The first experiment determines the algorithm's ability to differentiate between different instantaneous phases for the classification in the later steps. The second experiment investigates the performance of the technique with regards to the Doppler effect that causes changes in instantaneous frequency.
\begin{figure}[h!]
\centering
    \begin{subfigure}[b]{0.25\textwidth}
    \centering
        \includegraphics[width=\textwidth]{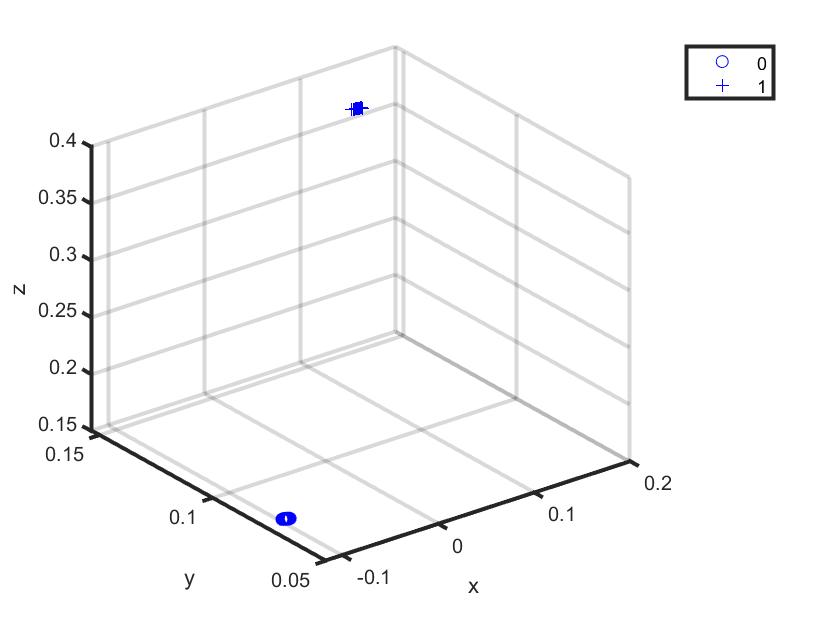}
        \caption{Feature for 2-PSK}
        \label{Feature for BPSK Modulation}
    \end{subfigure}
    \begin{subfigure}[b]{0.22\textwidth}
    \centering
        \includegraphics[width=\textwidth]{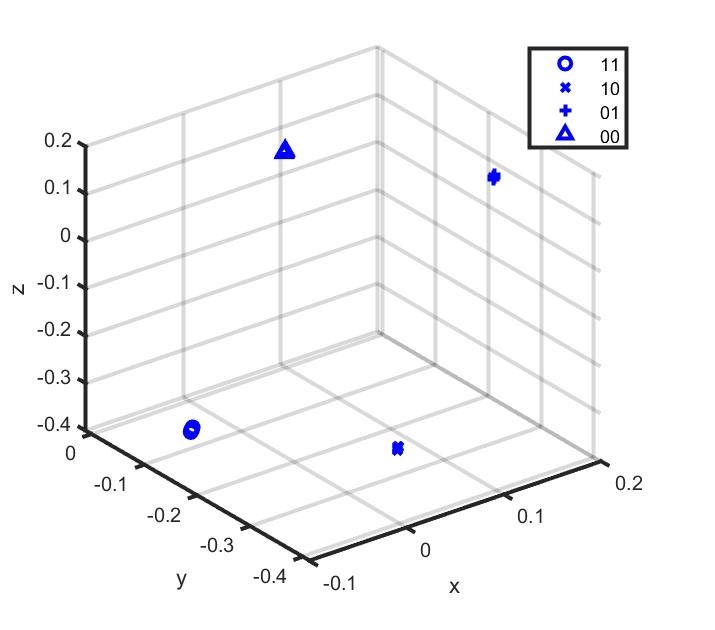}
        \caption{Feature for 4-PSK}
        \label{Feature for QPSK Modulation}
    \end{subfigure}\\
    \begin{subfigure}[b]{0.32\textwidth}
    \centering
        \includegraphics[width=\textwidth]{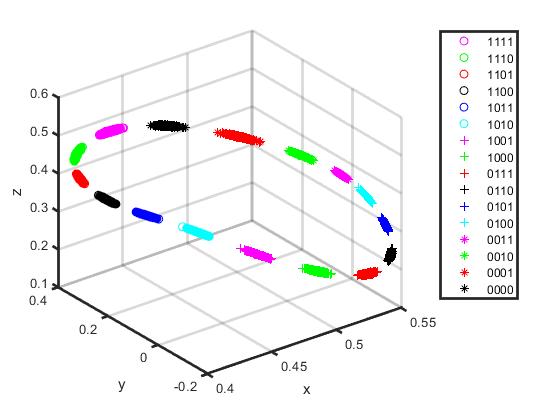}
        \caption{Feature for 16-PSK}
        \label{Feature for 16-PSK Modulation}
    \end{subfigure}
    \caption{First three latent DBN Features for 2-,4-,16-PSK Modulation Schemes}
    \label{Features plot}
\end{figure}

In Fig.~\ref{Features plot}, the first 3 latent features of the (28$\times$ 28) matrix $F(n)$ modelled individually for 2-, 4-, 16-PSK modulation schemes are shown. $F(n)$ depicts a strong similarity to the conventional PSK constellation, where each symbol is represented by a point on a 2-D axis.

\begin{figure}[h!]
\centering
    \begin{subfigure}[h]{0.36\textwidth}
    \centering
        \includegraphics[width=\textwidth]{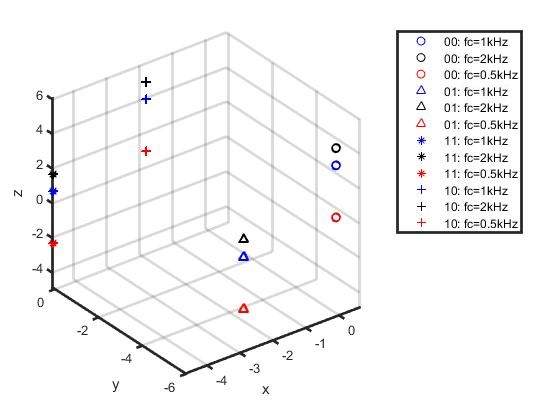}
        \caption{3-D Feature for 4-PSK}
        \label{Feature for 4-PSK}
    \end{subfigure}\\
    \begin{subfigure}[h]{0.36\textwidth}
    \centering
        \includegraphics[width=\textwidth]{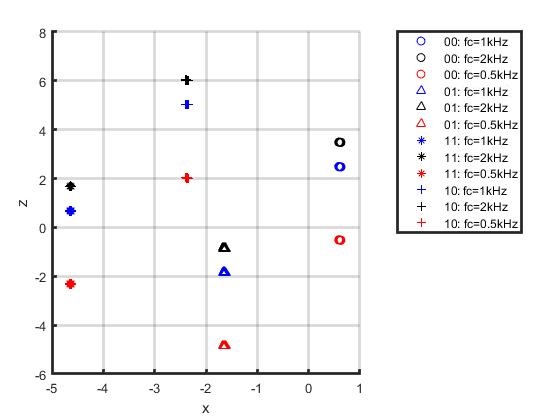}
        \caption{4-PSK xz-axis Feature}
        \label{4-PSK XZ-Plane Feature}
    \end{subfigure}%
    \caption{First three latent DBN Features for 4-PSK Modulation Schemes when $f_c$=0.5kHz,1kHz,2kHz, $f_s$=100Hz, bit rate = $0.05$ bits per sec}
    \label{DBN Features}
\end{figure}

Secondly, to evaluate the performance of the feature extraction with regards to different frequencies, we extracted features of signals of $f_c$=0.5kHz, $f_c$=1kHz, and $f_c$=2kHz respectively. Fig.~\ref{DBN Features} indicates that for 4-PSK, there is no discerning change on the xy plane. However, when the features are viewed on the xz plane, the features display a shift in the z-axis, which corresponds to the third latent feature.

\subsection{Simulation Results}
\label{Sim Results}

In this section, three experiments were conducted with different objectives. The first experiment assesses the performance of the two proposed algorithms by comparing their results with the conventional MLE. The second experiment evaluates the invariance of the algorithms in relation to the Doppler effect. The last experiment illustrate the accuracy of the algorithm with respects to the number of training data used. 

Firstly, to analyze the performance of the demodulation technique as a whole, the model is first evaluated against the conventional MLE, as shown in Fig.~\ref{BER for 2,4,8-PSK}. The generated dataset comprises of $s(t)=x(t)+n(t)$ and their binary representatives $y(n)$. 
\begin{figure}[hbt!]
\centering
\includegraphics[width=8.5cm]{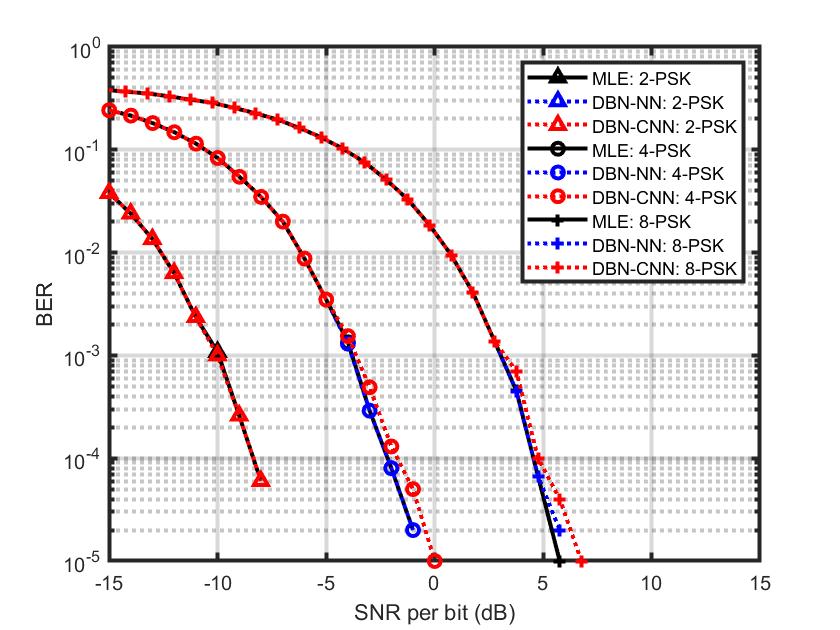}
\caption{BER comparison between DBN-CNN, DBN-NN and MLE for 2, 4, 8-PSK when $f_s$=100Hz, $f_c$=1kHz, bit rate = $0.05$ bits per sec}
\label{BER for 2,4,8-PSK}
\end{figure}
Both the proposed DBN-CNN and DBN-NN combinations achieves similar bit error rate (BER) as the MLE. However, for the 4- and 8-PSK, the proposed DBN-CNN shows an increase in error starting from signal-to-noise (SNR) at -4dB and 3dB respectively. There was a maximum of 0.5 dB error. 

Secondly, to investigate the Doppler invariant property of the two proposed algorithms, we analyzed the algorithms under different instantaneous frequencies. As the MLE is vulnerable to the Doppler effect and to demonstrate the performance of the feature extraction, this experiment will be evaluated against the MLE demodulation of signals without the Doppler effect. 

\begin{figure}[hbt!] %!t
\centering
\includegraphics[width=8.5cm]{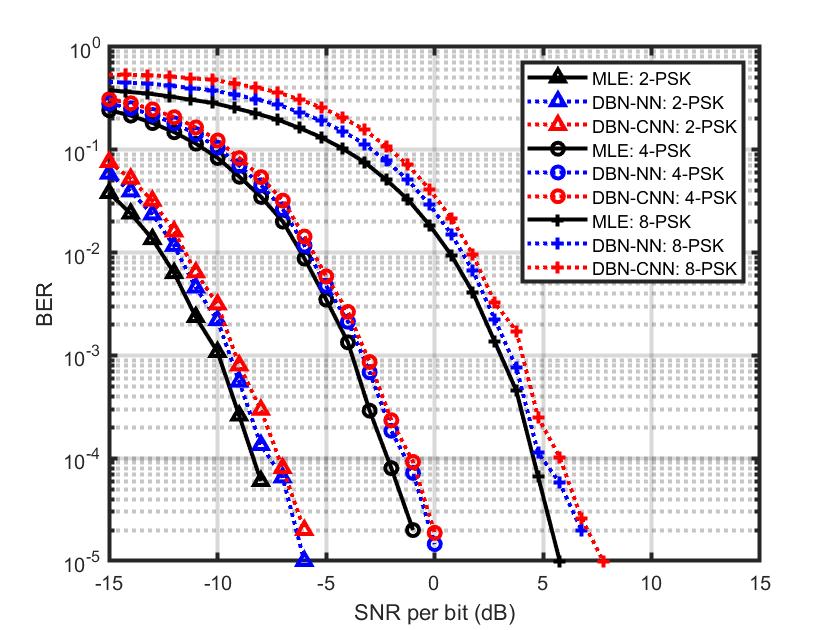}
\caption{BER comparison between DBN-CNN, DBN-NN and MLE for 2-, 4-, 8-PSK with varying Doppler co-efficients (To evaluate the Doppler invariant property of the proposed methods, MLE demodulation input signals are without the Doppler effect)}
\label{Varying freq}
\end{figure}

In this experiment, the channel model Eqn.\ref{channel} was used and the carrier frequency $f_c$ was randomly varied between 0.5kHz and 2kHz, using a normal distribution with a mean at $f_c$=1kHz and a standard deviation of 1. Fig.~\ref{Varying freq} presents the results of this experiment. 
Using the MLE as a baseline, this experiment illustrates that despite the varying Doppler effect, our algorithm is able to somewhat maintain the BER compared to the results shown in Fig.~\ref{BER for 2,4,8-PSK}. For 2-, 4- and 8-PSK, the increase in BER in comparison to the BER shown in Fig.~\ref{BER for 2,4,8-PSK} are relatively similar at 2dB. A reason for this is that the error occurred at the feature extraction phase. More research will have to be done to completely understand where the error was incurred. 

Both experiments confirmed the superior performance of the DBN-NN relative to the DBN-CNN. One of the possible reasons for this discrepancy is the extra layer that the CNN comprises of. This layer further extracts features from the feature image by controlling the resolution. However, this further defining of the features could result in over-extraction and thus, a higher BER. 
\begin{figure}[hbt!]
\centering
\includegraphics[width=7cm]{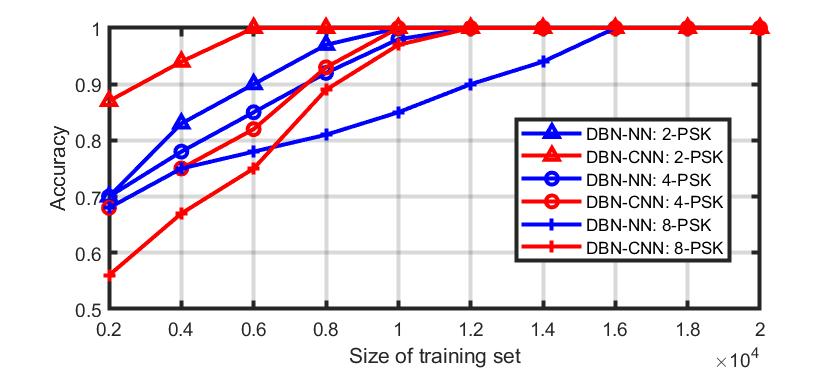}
\caption{Demodulation accuracy of 2-, 4-, 8-PSK modulated signals against the number of training periods}
\label{demodulation accuracy}
\end{figure}
In the final experiment, both the accuracy of the proposed DBN-NN and DBN-CNN are evaluated and shown in Fig\ref{demodulation accuracy}. The generated dataset consists of $x(t)$ and $y(t)$. It can be seen that the DBN-CNN reaches maximum accuracy with a smaller training set in comparison to DBN-NN. This might validate that the extra feature extraction layer in the CNN extracts more features from the feature image.
% ==================
% # Conclusion #
% ==================
\section{Conclusion}
\label{conclusion}
In this paper, we proposes two novel ML based approaches, DBN-NN and DBN-CNN, to the demodulation problem for SWAC channel's changes in instantaneous frequency. The proposed architecture consists of two parts --- a feature extraction technique, followed by a classification method. The framed received signal is inputted into the feature learning method and outputted as a feature image matrix. Next, this feature image is inputted to the classification models for labelling. The classification techniques predict the binary representation for each feature image via different NN layers.

The proposed systems were evaluated in three experiments and was found to be effective for the SWAC channel. As future work, the effectiveness and flexibility of the algorithms should be further explored. One possible area to investigate in future research is if the algorithm is able to differentiate more modulation schemes.

% ==================
% # Acknowledgment #
%\section*{Acknowledgment}
%For the Summary paper submission only, no %acknowledgements are allowed. 

% ==============
% # REFERENCES #
% ==============
\bibliographystyle{IEEEtran}
\bibliography{Demod}

\begin{thebibliography}{10}
\providecommand{\url}[1]{#1}
\csname url@samestyle\endcsname
\providecommand{\newblock}{\relax}
\providecommand{\bibinfo}[2]{#2}
\providecommand{\BIBentrySTDinterwordspacing}{\spaceskip=0pt\relax}
\providecommand{\BIBentryALTinterwordstretchfactor}{4}
\providecommand{\BIBentryALTinterwordspacing}{\spaceskip=\fontdimen2\font plus
\BIBentryALTinterwordstretchfactor\fontdimen3\font minus
  \fontdimen4\font\relax}
\providecommand{\BIBforeignlanguage}[2]{{%
\expandafter\ifx\csname l@#1\endcsname\relax
\typeout{** WARNING: IEEEtran.bst: No hyphenation pattern has been}%
\typeout{** loaded for the language `#1'. Using the pattern for}%
\typeout{** the default language instead.}%
\else
\language=\csname l@#1\endcsname
\fi
#2}}
\providecommand{\BIBdecl}{\relax}
\BIBdecl

\bibitem{A.C.Singer}
A.~Singer, J.~Nelson, and S.~Kozat, ``\BIBforeignlanguage{English (US)}{Signal
  processing for underwater acoustic communications},''
  \emph{\BIBforeignlanguage{English (US)}{IEEE Communications Magazine}},
  vol.~47, no.~1, pp. 90--96, 1 2009.

\bibitem{K.C.H.}
K.~C.~H. {Blom}, H.~S. {Dol}, A.~B.~J. {Kokkeler}, and G.~J.~M. {Smit}, ``Blind
  equalization of underwater acoustic channels using implicit higher-order
  statistics,'' in \emph{2016 IEEE Third Underwater Communications and
  Networking Conference (UComms)}, Aug 2016, pp. 1--5.

\bibitem{Zheng2010}
Z.~{Zheng}, ``Analysis and comparison of dimensional reduction based on capture
  data,'' in \emph{2010 Asia-Pacific Conference on Wearable Computing Systems},
  April 2010, pp. 163--164.

\bibitem{Mashford}
J.~S. {Mashford}, ``A neural network image classification system for automatic
  inspection,'' in \emph{Proceedings of ICNN'95 - International Conference on
  Neural Networks}, vol.~2, Nov 1995, pp. 713--717 vol.2.

\bibitem{Young}
T.~{Young}, D.~{Hazarika}, S.~{Poria}, and E.~{Cambria}, ``Recent trends in
  deep learning based natural language processing [review article],''
  \emph{IEEE Computational Intelligence Magazine}, vol.~13, no.~3, pp. 55--75,
  Aug 2018.

\bibitem{Wang}
Y.~Wang, H.~Zhang, Z.~Sang, L.~Xu, C.~Cao, and T.~A. Gulliver, ``Modulation
  classification of underwater communication with deep learning network,''
  \emph{Computational Intelligence and Neuroscience}, vol. 2019, pp. 1--12, 04
  2019.

\bibitem{Friedlander}
B.~{Friedlander} and J.~M. {Francos}, ``Estimation of amplitude and phase
  parameters of multicomponent signals,'' \emph{IEEE Transactions on Signal
  Processing}, vol.~43, no.~4, pp. 917--926, April 1995.

\bibitem{Stojanovic.M}
M.~Stojanovic, J.~Catipovic, and J.~G. Proakis, ``Adaptive multichannel
  combining and equalization for underwater acoustic communications,''
  \emph{The Journal of the Acoustical Society of America}, vol.~94, no.~3, pp.
  1621--1631, 1993.

\bibitem{Stojanovic1994}
M.~Stojanovic, J.~A. Catipovic, and J.~G. Proakis, ``Phase-coherent digital
  communications for underwater acoustic channels,'' \emph{IEEE Journal of
  Oceanic Engineering}, vol.~19, no.~1, pp. 100--111, Jan 1994.

\bibitem{Salama}
M.~A. {Salama}, A.~E. {Hassanien}, and A.~A. {Fahmy}, ``Deep belief network for
  clustering and classification of a continuous data,'' in \emph{The 10th IEEE
  International Symposium on Signal Processing and Information Technology}, Dec
  2010, pp. 473--477.

\bibitem{Stocia1989}
P.~{Stoica}, R.~L. {Moses}, B.~{Friedlander}, and T.~{Soderstrom}, ``Maximum
  likelihood estimation of the parameters of multiple sinusoids from noisy
  measurements,'' \emph{IEEE Transactions on Acoustics, Speech, and Signal
  Processing}, vol.~37, no.~3, pp. 378--392, March 1989.

\end{thebibliography}

\end{document}